\documentclass[12pt,]{article}

 \usepackage{graphicx}
 \usepackage[cp1251]{inputenc}  %% Windows

\begin{document}
%\leftline {УДК 524.6--34}\bigskip

 \centerline {\bf\Large Analysis of Galactic Rotation from Masers }
 \centerline {\bf\Large Based on a Nonlinear Oort Model}
 \bigskip
 \centerline
 {
  V.V. Bobylev$^{1,2}$
\footnote [0]{e-mail: vbobylev@gao.spb.ru},
 A.T.Bajkova$^1$
 }
 \bigskip
 \centerline {\small $^{1}$\it Pulkovo Astronomical Observatory, Russian Academy of
Sciences,} \centerline {\small\it Pulkovskoe sh. 65, St.
Petersburg, 196140 Russia}
 \centerline {\small $^{2}$\it Sobolev Astronomical Institute, St. Petersburg State University,
}
 \centerline {\small \it Universitetskii pr. 28, Petrodvorets, 198504 Russia}
 \bigskip

{\noindent Based on data on Galactic masers with measured
trigonometric parallaxes, we have tested a nonlinear model of
Galactic rotation using generalized Oort formulas. This model is
shown to yield pretty good results up to heliocentric distances of
3--4~kpc. The main feature of the method is the possibility of
estimating the solar Galactocentric distance $R_0$. This distance
has been found by analyzing $\approx$60 masers to be
$R_0=8.3\pm0.4$ kpc. Our study of the three-dimensional kinematics
of more than 100 masers based on the Ogorodnikov--Milne model has
shown that significant nonlinearities are present only in the $xy$
plane (rotation around the Galactic $z$ axis) due to the
peculiarities of the Galactic rotation curve. No significant
linear dependences have been found in the $xz$ and $yz$ planes. We
show the presence of a wave in the velocities $w$ as a function of
coordinate $x$ or distance $R$ with a wavelength of 3 kpc and an
amplitude of 10 km s$^{-1}$. The wave is particularly prominent in
the Local and Perseus arms.}

 \bigskip\noindent

\noindent{\bf DOI:} 10.1134/S1063773714120019

\medskip


\newpage
\section*{INTRODUCTION}

The parameters of Galactic rotation have been repeatedly
determined by many authors using objects belonging to various
structural components of the Galaxy: from ionized and neutral
hydrogen (Clemens 1985; McClure-Griffiths and Dickey 2007; Levine
et al. 2008), from distant OB associations (Mel’nik and Dambis
2009), and from open star clusters (Zabolotskikh et al. 2002;
Popova and Loktin 2005; Bobylev et al. 2007). Galactic masers are
of great interest for studying the Galactic kinematics, because
their trigonometric parallaxes are measured by VLBI with a high
accuracy, on average, at least 10\% (Reid et al. 2014).

One of the important problems is to determine the Galactocentric
distance of the Sun $R_0$. Reid (1993) published a review of the
R0 measurements made by then by various methods. He divided all
measurements into primary, secondary, and indirect ones and
obtained the “best value” as a weighted mean of the published
measurements over a period of 20 years: $R_0 = 8.0\pm 0.5$ kpc.
Nikiforov (2003) proposed a more complete three-dimensional
classification in which the type of R0 determination method, the
method of finding the reference distances, and the type of
reference objects are taken into account. Taking into account the
main types of errors and correlations associated with the classes
of measurements, he obtained the “best value” $R_0 = 7.9\pm 0.2$
kpc by analyzing the results of various authors published between
1974 and 2003. Based on the results published between 1992 and
2010, Foster and Cooper (2010) obtained the mean $R_0 = 8.0\pm
0.4$ kpc. Since the scatter of individual estimates is
significant, applying independent methods is of great interest.

The direct methods of determining the distance R0 are not all that
many. These include such methods as the VLBI measurements of the
trigonometric parallaxes for masers at the Galactic center, the
use of Cepheids and globular clusters, or the application of the
dynamical parallax method to analyze the motion of stars around
the supermassive central black hole. There are much more indirect
methods. The various kinematic methods of estimating $R_0$ occupy
an important place among them. In particular, using such a method,
Reid et al. (2014) obtained an estimate of $R_0 = 8.34\pm 0.16$
kpc by analyzing the kinematics of Galactic masers, while Bobylev
and Bajkova (2014) found $R_0 = 8.3\pm 0.2$ kpc from masers, with
these authors having applied different methods of analysis.

The number of masers with measured trigonometric parallaxes
already exceeds 100. This allows one not only to study the
Galactic rotation parameters and the influence of the spiral
structure (the $xy$ plane) but also to check the presence of some
kinematic peculiarities in the other two planes ($xz$ and $yz$).
Such a check is one of the goals of this paper. For example, a
kinematic relationship to the Galactic warp (Bobylev 2013), which
manifests itself in the $yz$ plane, is revealed in the motion of
distant Cepheids. Having analyzed OB stars, Branham (2014) found
significantly nonzero values of the gradients $\partial w/\partial
z$ and $\partial^2 w/\partial^2 z$. He used a nonlinear model
based on generalized Oort formulas, which served as a stimulus for
applying a similar approach in this paper. Vityazev and Tsvetkov
(2014) reported the existence of significantly nonzero
beyond-the-model (with respect to the linear Ogorodnikov.Milne
model) harmonics in the motion of stars from various catalogs.

The main goal of this paper is to test a nonlinear model of
Galactic rotation using generalized Oort formulas. Such an
approach allows the angular velocity of Galactic rotation and its
derivatives as well as the solar Galactocentric distance $R_0$ to
be estimated.

\section*{DATA}
Based on published data, we gathered information about the
coordinates, line-of-sight velocities, proper motions, and
trigonometric parallaxes of Galactic masers measured by VLBI with
an error, on average, less than 10\%. These masers are associated
with very young objects, protostars of mostly high masses located
in regions of active star formation. The proper motions and
trigonometric parallaxes of the masers are absolute, because they
are determined with respect to extragalactic reference objects
(quasars).

One of the projects to measure the trigonometric parallaxes and
proper motions is the Japanese VERA (VLBI Exploration of Radio
Astrometry) project devoted to the observations of H$_2$O masers
at 22.2 GHz (Hirota et al. 2007) and a number of SiO masers (which
are very few among young objects) at 43 GHz (Kim et al. 2008).

Methanol (CH$_3$OH, 6.7 and 12.2 GHz) and H$_2$O masers are
observed in the USA on VLBA (Reid et al. 2009a). Similar
observations are also being carried out within the framework of
the European VLBI network (Rygl et al. 2010), in which three
Russian antennas are involved: Svetloe, Zelenchukskaya, and
Badary. These two programs enter into the BeSSeL project$^1$ (Bar
and Spiral Structure Legacy Survey, Brunthaler et al. 2011).
 \footnote {http://www3.mpifr-bonn.mpg.de/staff/abrunthaler/BeSSeL/index.shtml}

The VLBI observations of radio stars in continuum at 8.4 GHz are
being carried out with the same goals (Torres et al. 2009; Dzib et
al. 2011). Radio sources located in the local (Orion) arm
associated with young low-mass protostars are observed within the
framework of this program.

Our sample includes a total of 117 sources. In addition to 103
masers from Reid et al. (2014), it includes 14 more sources whose
trigonometric parallaxes have been measured by VLBI:

(1) the high-mass spectroscopic binary Cyg X-1 (Reid et al. 2011),
with one of its components being a black hole candidate, we took
the line-of-sight velocity for this binary $V_\gamma$ from
Ziolkowski (2005);

(2) the red supergiant IRAS 22480+6002 (Imai et al. 2012);

(3) the red supergiant PZ~Cas (Kusuno et al. 2013);

(4) the source SVS 13 in NGC~1333 (Hirota et al. 2008), with the
data having been averaged over two features, f1 and f2;

(5–9) five low-mass nearby radio stars in Taurus: Hubble 4 and
HDE~283572 (Torres et al. 2007), T~Tau~N (Loinard et al. 2007),
HP~Tau/G2 (Torres et al. 2009), and V773~Tau (Torres et al. 2012);

(10–12) three radio stars: S1 and DoAr21 (Loinard et al. 2008) as
well as EC~95 (Dzib et al. 2010) with their line-of-sight
velocities $V_{LSR}$ from Honma et al. (2012);

(13) IRAS 20143+3634 (Burns et al. 2014);

(14) IRAS 22555+6213 (Chibueze et al. 2014).

Note that we took a new estimate of the distance and proper motion
for the star-forming region S 269 (which enters into the sample of
103 masers) from Asaki et al. (2014).

\section*{METHODS}
Here, we use a rectangular Galactic coordinate system with the
axes directed away from the observer toward the Galactic center
$(l$=$0^\circ$, $b$=$0^\circ,$ the $x$ axis), in the direction of
Galactic rotation $(l$=$90^\circ$,~$b$=$0^\circ,$ the $y$ axis),
and toward the north Galactic pole $(b$=$90^\circ,$ the $z$ axis).

\subsection*{Generalized Oort Equations}
Bottlinger’s equations are
 \begin{equation}
 \begin{array}{lll}
 V_r=-(\omega-\omega_0)R_0\sin l \cos b,\\
 V_l=-(\omega-\omega_0)R_0\cos l+\omega r\cos b,\\
 V_b= (\omega-\omega_0)R_0\sin l \sin b,
 \label{Bottl-0234}
 \end{array}
 \end{equation}
where the signs at the angular velocities of Galactic rotation
$\omega$ and $\omega_0$ correspond to positive rotation from the
$x$ axis to $y$, i.e., counterclockwise rotation. The angular
velocity $\omega$ is then expanded in a series to terms of the
second order in $r/R_0$. In this case, the distance from the star
to the Galactic rotation axis $R$,
 \begin{equation}
 \begin{array}{lll}
 R^2=r^2\cos^2 b-\\
    -2R_0 r\cos b\cos l+R^2_0
 \label{RR}
 \end{array}
 \end{equation}
is again expanded in a series in $r/R_0$ and then only the first
two expansion terms are used: $R\approx R_0-r\cos b\cos l.$ The
resulting expressions are called the generalized Oort equations.
They are written as (Ogorodnikov 1965)
 \begin{equation}
 \begin{array}{lllll}
  V_r=-u_\odot\cos b\cos l-v_\odot\cos b\sin l-\\
    -w_\odot\sin b+rA\cos^2 b\sin 2l-\\
    -r^2F\cos^3 b\sin l\cos^2 l-\\
    -r^2(A^\circ/R_0)\cos^3 b\sin^3 l,
 \label{Ogor-01}
 \end{array}
 \end{equation}
 \begin{equation}
 \begin{array}{lllll}
 V_l=u_\odot\sin l-v_\odot\cos l+\\
  +rA\cos b\cos 2l+ rB\cos b-\\
  -r^2F\cos^2 b\cos^3 l-\\
  -r^2(A^\circ/R_0)(3\cos^2 b\cos l-\\
  -\cos^2 b\cos^3 l),
 \label{Ogor-02}
 \end{array}
 \end{equation}
 \begin{equation}
 \begin{array}{lllll}
 V_b=u_\odot\cos l\sin b+v_\odot\sin l\sin b-\\
 -w_\odot\cos b-rA\cos b\sin b\sin 2l+\\
 +r^2F\cos^2 b\sin b\sin l\cos^2 l+\\
 +r^2(A^\circ/R_0)\cos^2 b\sin b\sin^3 l,
 \label{Ogor-03}
 \end{array}
 \end{equation}
where $V_r$ is the line-of-sight velocity of the star (in km
s$^{-1}$); $r=1/\pi$ is the heliocentric distance of the
star;$V_l=4.74 r \mu_l\cos b$ and $V_b=4.74 r \mu_b$ are the
proper motion velocity components of the star (in mas yr$^{-1}$)
in the $l$ and $b$ directions, respectively; the coefficient
$4.74$ is the quotient of the number of kilometers in an
astronomical unit by the number of seconds in a tropical year;
$u_\odot,v_\odot,w_\odot$ are the stellar group velocity
components relative to the Sun taken with the opposite sign (the
velocity $u$ is directed toward the Galactic center, $v$ is in the
direction of Galactic rotation, $w$ is directed to the north
Galactic pole); $R_0$ is the Galactocentric distance of the Sun;
$\omega_0$ is the angular velocity of rotation at the distance
$R_0$; the parameters $\omega'_0$ and $\omega''_0$ are the first
and second derivatives of the angular velocity, respectively

The system of conditional equations
(\ref{Ogor-01})--(\ref{Ogor-03}) contains seven unknowns:
$u_\odot$,$v_\odot$,$w_\odot$,
 $A$, $B$, $F$ and $A^\circ/R_0$, which are determined by the
least-squares method. Here, $A$ and $B$ denote the Oort constants,
$A=0.5R_0\omega'_0$ and $B=0.5R_0\omega'_0+\omega_0$,
consequently, $\omega_0=B-A$, while $F$ denotes the second-order
coefficient $F=0.5R_0\omega''_0$. $R_0$ is calculated using the
values of $A$ and $A^\circ/R_0$ found as $R_0=A/(A^\circ/R_0)$.
The error in $R_0$ is found from the relation
$\varepsilon(R_0)=\varepsilon(A)/(A^\circ/R_0)$.

As a result, in addition to the peculiar solar velocity components
 $u_\odot$,$v_\odot$,$w_\odot$, the method allows the Galactic rotation parameters
$\omega_0$, $\omega'_0$, and $\omega''_0$  and the distance $R_0$
to be estimated irrespective of any a priori assumptions.

\subsection*{The Linear Ogorodnikov–Milne Model}
In the linear Ogorodnikov–Milne model (Ogorodnikov 1965), the
observed velocity ${\bf V}(r)$ of a star with a heliocentric
radius vector ${\bf r}$ is described, to terms of the first order
of smallness $r/R_0\ll 1$, by the vector equation
\begin{equation}
 {\bf V}(r)={\bf V}_\odot+M{\bf r}+{\bf V'}.
 \label{eq-1}
 \end{equation}
Here, ${\bf V}_\odot(u_\odot,v_\odot,w_\odot)$ is the Sun’s
peculiar velocity relative to the stars under consideration, {$\bf
V'$} is the star’s residual velocity, $M$ is the displacement
matrix whose components are the partial derivatives of the
velocity ${\bf u}(u,v,w)$ with respect to the distance ${\bf
r}(x,y,z)$, where ${\bf u}={\bf V}(R)-{\bf V}(R_0)$. Neglecting
the residual velocities, Eqs. (\ref{eq-1}) can be written in an
expanded form as
 \renewcommand{\arraystretch}{2.4}
 \arraycolsep=1pt
 \begin{equation}
 \begin{array}{rllcc}
 u&=&u_\odot+{\displaystyle \left(\frac{\partial u}{\partial x}\right)}_\circ x+
             {\displaystyle \left(\frac{\partial u}{\partial y}\right)}_\circ y+
             {\displaystyle \left(\frac{\partial u}{\partial z}\right)}_\circ z, &&\\
 v&=&v_\odot+{\displaystyle \left(\frac{\partial v}{\partial x}\right)}_\circ x+
             {\displaystyle \left(\frac{\partial v}{\partial y}\right)}_\circ y+
             {\displaystyle \left(\frac{\partial v}{\partial z}\right)}_\circ z, &&\\
 w&=&w_\odot+{\displaystyle \left(\frac{\partial w}{\partial x}\right)}_\circ x+
             {\displaystyle \left(\frac{\partial w}{\partial y}\right)}_\circ y+
             {\displaystyle \left(\frac{\partial w}{\partial z}\right)}_\circ z, &&
 \label{OG-M-99} \end{array}\end{equation}
\renewcommand{\arraystretch}{1.2}
where the subscript 0 means that the derivatives are taken at
$R=R_0.$ In cylindrical coordinates $(R,\theta,z)$, the
displacement matrix $M$ is
 \begin{equation}
     \displaystyle
 M=\pmatrix {
  {\strut\displaystyle \partial\displaystyle V_R\over\displaystyle\partial\displaystyle R}&
 ~~{\strut\displaystyle 1\over\displaystyle R}{\strut\displaystyle\partial\displaystyle
 V_R\over\displaystyle\partial\displaystyle\theta}-{\strut\displaystyle  V_\theta\over\displaystyle R}~~&
  {\strut\displaystyle \partial\displaystyle  V_R\over\displaystyle\partial\displaystyle z}\cr
  {\strut\displaystyle \partial\displaystyle  V_\theta\over\displaystyle\partial\displaystyle  R}&
 ~~{\strut\displaystyle 1\over\displaystyle R}{\strut\displaystyle\partial\displaystyle
 V_\theta\over\displaystyle\partial\displaystyle\theta}+{\strut\displaystyle  V_R\over\displaystyle R}~~&
  {\strut\displaystyle \partial\displaystyle  V_\theta\over\displaystyle\partial\displaystyle z}\cr
  {\strut\displaystyle \partial\displaystyle  V_z\over\displaystyle\partial\displaystyle  R}&
  {\strut\displaystyle 1\over\displaystyle R}{\strut \displaystyle \partial\displaystyle V_z\over\displaystyle\partial\displaystyle \theta}&
  {\strut\displaystyle \partial\displaystyle  V_z\over\displaystyle\partial\displaystyle  z}\cr
 }.
 \label{OG-M-99-cil} \end{equation}
Here, all derivatives are taken at the point with coordinates
$(R=R_\circ,\theta=0^\circ,z=z_0)$.

\section*{RESULTS}
Following the recommendations of Reid et al. (2014), we did not
use any masers located at distances $R<4$ kpc when determining the
Galactic rotation parameters, because the influence of the
noncircular motions associated with the influence of the central
Galactic bar is great in this region. To determine the Galactic
rotation parameters, we produced a sample of ``best'' masers. For
this purpose, we solved the system of equations
~(\ref{Ogor-01})--(\ref{Ogor-03}) using the constraints
$\varepsilon_\pi/\pi\leq10\%$ and $r\leq4$ kpc that are met by 66
masers. As a result of successive rejections according to the
$3\sigma$ criterion, 60 sources remained. The following masers
turned out to be the rejected ones: G012.68.00.18, G016.58.00.05,
G034.39+00.22, G078.12+03.63, G078.88+00.70, and G108.59+00.49. As
a result, we obtained the following solution:
 \begin{equation}
 \begin{array}{lll}
  u_\odot= 8.5\pm1.3~\hbox{km s$^{-1}$}, \\
  v_\odot=15.7\pm1.4~\hbox{km s$^{-1}$}, \\
  w_\odot= 9.0\pm1.0~\hbox{km s$^{-1}$}, \\
   A= 15.03\pm0.65~\hbox{km s$^{-1}$ kpc$^{-1}$}, \\
   B=-14.94\pm0.69~\hbox{km s$^{-1}$ kpc$^{-1}$}, \\
   F= -2.74\pm0.69~\hbox{km s$^{-1}$ kpc$^{-2}$}, \\
 A^\circ/R_0=1.80\pm0.25~\hbox{km s$^{-1}$ kpc$^{-2}$},
 \label{rez-3}
 \end{array}
 \end{equation}
based on which we found
 $$
 \begin{array}{lll}
   \omega_0= -29.97\pm0.95~\hbox{km s$^{-1}$ kpc$^{-1}$},    \\
  \omega'_0=  3.59\pm0.16~\hbox{km s$^{-1}$ kpc$^{-2}$},\\
 \omega''_0= -0.66\pm0.16~\hbox{km s$^{-1}$ kpc$^{-3}$},\\
   R_0= 8.35\pm0.36~\hbox{kpc}.
 \end{array}
 $$
The error per unit weight is $\sigma_0=7.9$ km s$^{-1}$, with
solution (\ref{rez-3}) having been obtained with unit weights. In
the case of applying a system of weights in the form $w_r=
S_0/\sqrt{S_0^2+\sigma^2_{V_r}},$
 $w_l= S_0/\sqrt {S_0^2+\sigma^2_{V_l}}$ and
 $w_b= S_0/\sqrt {S_0^2+\sigma^2_{V_b}},$
where $S_0$ denotes the “cosmic” dispersion taken to be 8 km
s$^{-1}$, the solution is
 \begin{equation}
 \begin{array}{lll}
  u_\odot= 9.5\pm1.3~\hbox{km s$^{-1}$}, \\
  v_\odot=15.9\pm1.4~\hbox{km s$^{-1}$}, \\
  w_\odot= 8.6\pm0.9~\hbox{km s$^{-1}$}, \\
   A= 15.64\pm0.66~\hbox{km s$^{-1}$ kpc$^{-1}$}, \\
   B=-15.29\pm0.73~\hbox{km s$^{-1}$ kpc$^{-1}$}, \\
   F= -3.16\pm0.79~\hbox{km s$^{-1}$ kpc$^{-2}$}, \\
 A^\circ/R_0=1.84\pm0.28~\hbox{km s$^{-1}$ kpc$^{-2}$},
 \label{rez-4}
 \end{array}
 \end{equation}
based on which we find
 $$
 \begin{array}{lll}
   \omega_0= -30.93\pm0.99~\hbox{km s$^{-1}$ kpc$^{-1}$},   \\
  \omega'_0=  3.68\pm0.16~\hbox{km s$^{-1}$ kpc$^{-2}$},\\
 \omega''_0= -0.74\pm0.19~\hbox{km s$^{-1}$ kpc$^{-3}$},\\
   R_0= 8.50\pm0.36~\hbox{kpc},
 \end{array}
 $$
with the error per unit weight being $\sigma_0=7.6$ km s$^{-1}$.
It can be seen that there are no significant differences from the
solution with unit weights in this case.

Below, we give one more solution obtained under the constraints
$\varepsilon_\pi/\pi\leq10\%$ and $r\leq3$ kpc (53 sources) using
the weights described above:
 \begin{equation}
 \begin{array}{lll}
  u_\odot=10.3\pm1.4~\hbox{km s$^{-1}$}, \\
  v_\odot=15.9\pm1.4~\hbox{km s$^{-1}$}, \\
  w_\odot= 8.1\pm1.0~\hbox{km s$^{-1}$}, \\
   A= 16.46\pm0.74~\hbox{km s$^{-1}$ kpc$^{-1}$}, \\
   B=-15.48\pm0.80~\hbox{km s$^{-1}$ kpc$^{-1}$}, \\
   F= -3.81\pm0.96~\hbox{km s$^{-1}$ kpc$^{-2}$}, \\
 A^\circ/R_0=1.97\pm0.36~\hbox{km s$^{-1}$ kpc$^{-2}$},
 \label{rez-55}
 \end{array}
 \end{equation}
then
 $$
 \begin{array}{lll}
   \omega_0= -31.94\pm1.09~\hbox{km s$^{-1}$ kpc$^{-1}$},   \\
  \omega'_0=  3.95\pm0.18~\hbox{km s$^{-1}$ kpc$^{-2}$},\\
 \omega''_0= -0.91\pm0.23~\hbox{km s$^{-1}$ kpc$^{-3}$},\\
   R_0= 8.33\pm0.38~\hbox{kpc},
 \end{array}
 $$
with the error per unit weight being $\sigma_0=7.2$ km s$^{-1}$.
In comparison with solutions ((\ref{rez-3}) and (\ref{rez-4}), the
values of such local parameters as $w_\odot,$ $A$ and $B,$ are
determined here slightly better, but the value of $\omega''_0$ is
determined less reliably because of the reduction in the sample
radius.

%%%%%%%%%%%%%%%%%%%%%%%%%%%%%%%%%%%%%%%% FIG.1:
 \begin{figure}[t]
 {\begin{center}
 \includegraphics[width=120mm]{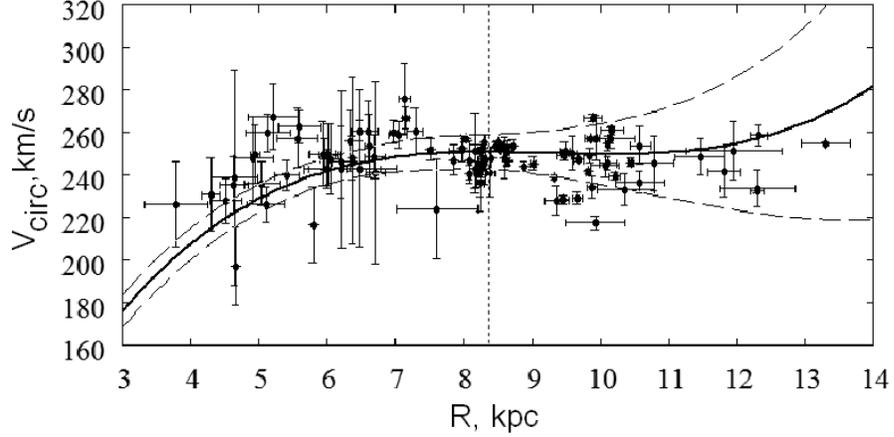}
 \caption{Galactic rotation curve constructed with parameters (\ref{rez-3}); the thin lines mark the $1\sigma$ confidence region, the vertical
dashed line marks the Sun’s position found. }
 \label{fig1}
 \end{center} }
 \end{figure}
%%%%%%%%%%%%%%%%%%%%%%%%%%%%%%%%%%%%%%%%%%%%%%%
%%%%%%%%%%%%%%%%%%%%%%%%%%%%%%%%%%%%%%%% FIG.2:
 \begin{figure}[t]
 {\begin{center}
 \includegraphics[width=160mm]{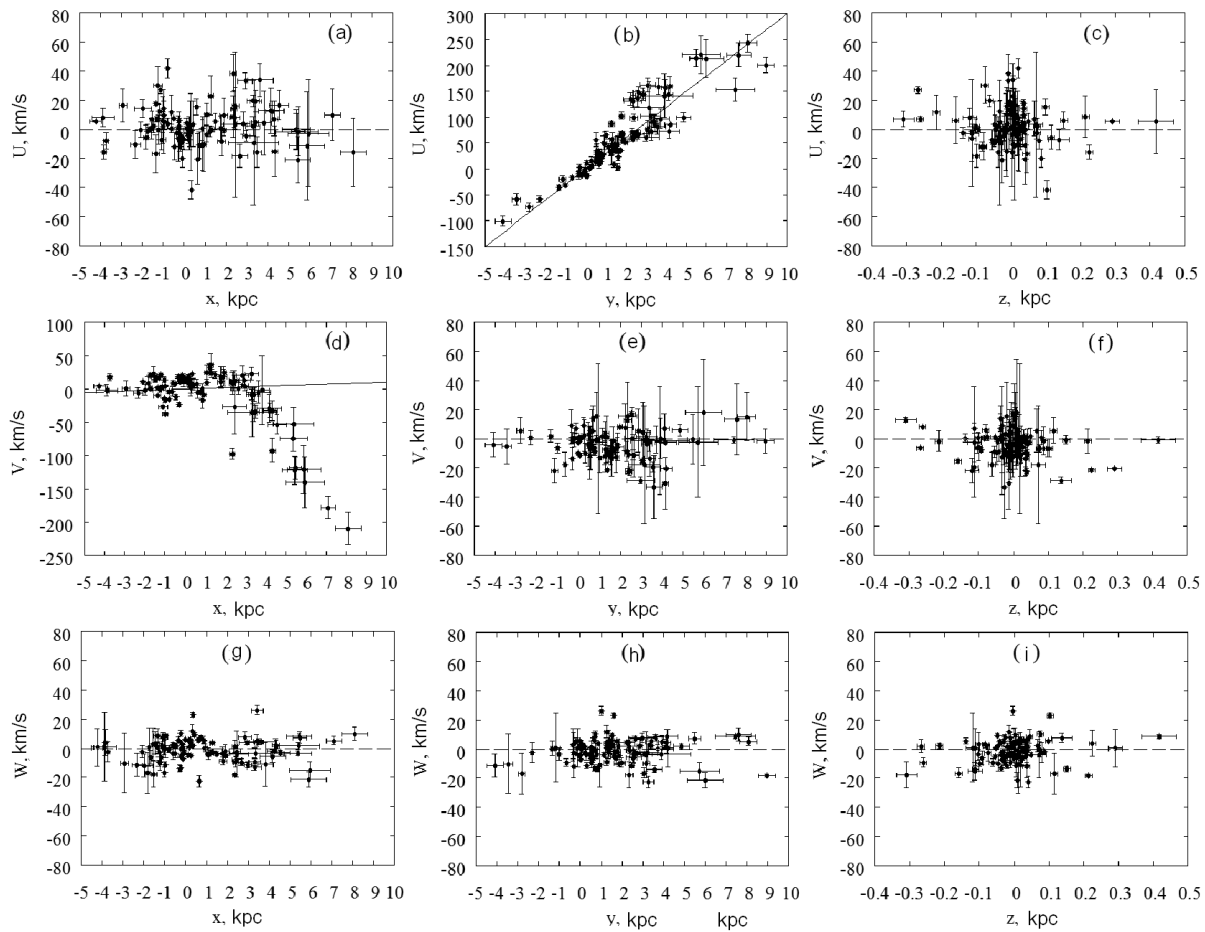}
 \caption{Velocities of the sample of 106 masers versus coordinates. The velocities on panels (b) and (d) are not the residual ones;
the velocities $u$ and $v$ on the other panels are the residual
ones. }
 \label{fig2}
 \end{center} }
 \end{figure}
%%%%%%%%%%%%%%%%%%%%%%%%%%%%%%%%%%%%%%%%%%%%%%%
%%%%%%%%%%%%%%%%%%%%%%%%%%%%%%%%%%%%%%%% FIG.3:
 \begin{figure}[t]
 {\begin{center}
 \includegraphics[width=120mm]{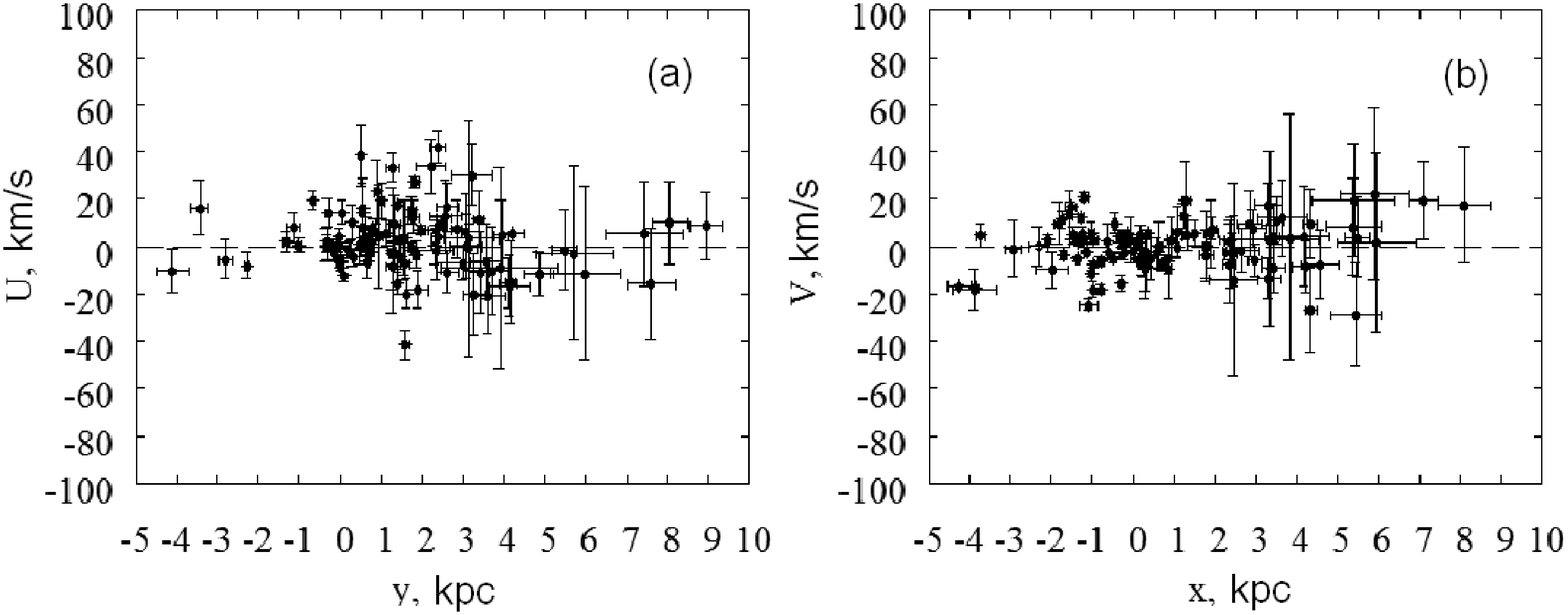}
 \caption{Residual velocities $u$ and $v$ of the masers versus coordinates y and x, respectively.}
 \label{fig3}
 \end{center} }
 \end{figure}
%%%%%%%%%%%%%%%%%%%%%%%%%%%%%%%%%%%%%%%%%%%%%%%

Figure ~\ref{fig1} shows an example of the Galactic rotation
curve, $V_{circ}=|R\omega|$, constructed with parameters
(\ref{rez-3}), where the velocities of 106 masers are plotted. As
can be seen from the figure, the rotation curve describes
satisfactorily the distribution of maser rotation velocities in
the range $4<R<12$ kpc.

It can be concluded that the method based on the generalized Oort
equations yields pretty good results and extends the capabilities
of the linear Oort–-Lindblad model (which can be used up to
distances of 2 kpc) to heliocentric distances of 3–-4 kpc, while
the main feature of the method is that the distance $R_0$ can be
estimated fairly easily. As we see from the solutions obtained
from masers, this distance is $R_0=8.3\pm0.4$ kpc.

Let us now turn to the Ogorodnikov–Milne model. Figure \ref{fig2}
presents nine dependences of the maser velocities $u,v,w$ on
coordinates $x,y,z$. In fact, all nine elements of the
displacement matrix $M$~(\ref{eq-1}) are displayed here
graphically, with the panels in the figure following in the same
order as the elements of the matrix M in forms (\ref{OG-M-99}) and
(\ref{OG-M-99-cil}).

The Galactic rotation around the $z$ axis is described by the
velocities $u$ and $v$ that lie in the Galactic $xy$ plane. Four
panels, (a), (b), (d), and (e), refer to this motion. The
velocities on panels (b) and (d) are not the residual ones, while
those on the remaining panels are the residual ones that were
formed by taking into account the Galactic rotation with
parameters (\ref{rez-3}) in the velocities $V_r,V_l$ and $V_b$.
The thin solid line on panel (b) plots the dependence ${\partial
u}/{\partial y}= 30$ km s$^{-1}$ kpc$^{-1}$; this is the angular
velocity of Galactic rotation taken with the opposite sign,
$-\omega_0=-(B-A)=-{\strut\displaystyle V_\theta/\displaystyle
R}$. The plotted linear dependence is seen to represent
excellently the data in the wide range of the coordinate
$y=[-3;8]$ kpc. This suggests that the term
${\strut\displaystyle\partial\displaystyle
 V_R/(R\displaystyle\partial\displaystyle\theta})$ in matrix
(\ref{OG-M-99-cil}) may be set equal to zero. Since there are no
significant linear trends on panels (a) and (e), it can be
concluded that no large-scale effects like expansion/contraction
are present in the Galaxy in the range of distances under
consideration. The thin solid line on panel (d) plots the
dependence ${\partial v}/{\partial x}= 1$ km s$^{-1}$ kpc$^{-1}$;
this is the first derivative of the linear Galactic rotation
velocity, $A+B= {\partial\displaystyle
V_\theta/\partial\displaystyle R}$. The application of the linear
approach is clearly seen to be limited by a radius of $\approx 2$
kpc from the Sun; the velocities v beyond this range deviate
significantly from the linear trend.

Figure \ref{fig3} presents the dependences of the residual maser
velocities $u$ and $v$ on coordinates $y$ and $x$, respectively.
Allowance for the Galactic rotation parameters (\ref{rez-3}) found
with two derivatives of the angular velocity is seen to correct
the data quite well, although a wave with a small amplitude
($\approx$10 km s$^{-1}$) is visible in the velocities u on the
left panel of Fig.\ref{fig3}.

Panels (e), (f), (h), and (i) in Fig.\ref{fig2} refer to the
motions in the $yz$ plane. As we see from the figure, no
significant linear trends are traceable here. This plane is
interesting in that the velocities of the objects can have a
kinematic relationship to such a large scale phenomenon in the
Galaxy as the disk warp (Bobylev 2013). The masers at great (more
than 10 kpc) heliocentric distances in directions
$l\approx90^\circ$ and $l\approx270^\circ$ with considerable $z$
are clearly not yet enough to study this effect based on them.

Panels (a), (c), (g), and (i) in Fig.\ref{fig2} refer to the
motions in the $xz$ plane. No significant linear trends are
traceable here either. Panel (g), where a periodicity with a
wavelength of 3--4 kpc and an amplitude of $\approx 10$ km
s$^{-1}$ is clearly seen, engages our attention. According to the
linear model of the spiral structure by Lin and Shu (1964), only
the velocities $V_R$ and $V_\theta$ are subjected to perturbations
from the density wave. Manifestations of these perturbations are
clearly seen on panels (a) and (d) in Fig.\ref{fig2} as well as on
the right panel in Fig.\ref{fig3}.

\begin{figure}[t]
 {\begin{center}
 \includegraphics[width=100mm]{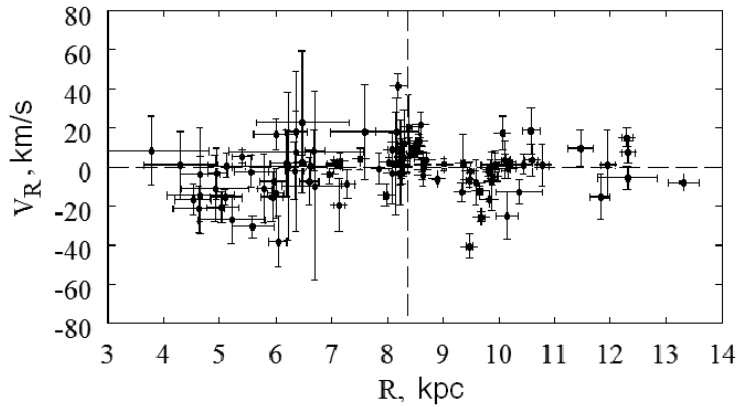}
 \caption{
Radial velocity $V_R$ versus Galactocentric distance $R$; the
vertical dashed line marks the Sun’s position found. }
 \label{fig4}
 \end{center} }
 \end{figure}

The tangential velocity perturbations produced by the spiral
density wave are clearly seen in Fig.\ref{fig1} as periodic
deviations from the smooth rotation curve. They manifest
themselves even better in the radial velocities $V_R,$, which can
be easily seen from Fig.\ref{fig4}.

\begin{figure}[t]
 {\begin{center}
 \includegraphics[width=140mm]{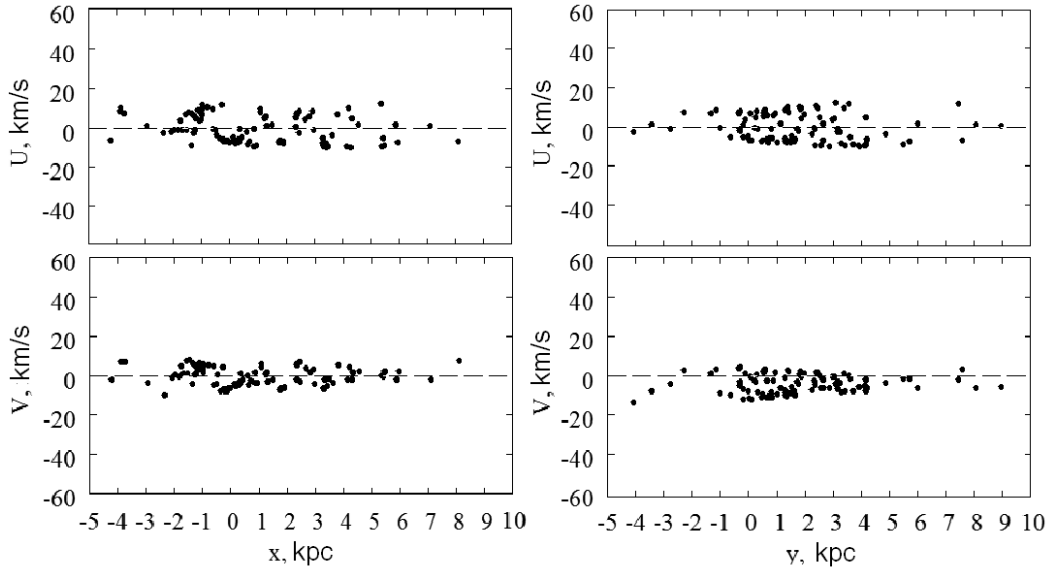}
 \caption{
Example of the influence of the spiral structure on the velocity
components $u$ and $v$ as a function of coordinates $x$ and $y$;
for details, see the text. }
 \label{fig5}
 \end{center} }
 \end{figure}

Figure \ref{fig5} shows an example of the influence of the spiral
structure on the velocity components $u$ and $v$. The corrections
were calculated for each maser with the following model
parameters: equal (in magnitude) amplitudes, $f_R=-8$ km s$^{-1}$
and $f_\theta=8$ km s$^{-1}$1, but individual phases of the Sun in
the wave, $(\chi_\odot)_R= -160^\circ$ and
$(\chi_\odot)_\theta=-50^\circ,$ the wavelength $\lambda=2.4$ kpc,
a two-armed spiral pattern, $m=2$. These parameters of the spiral
density wave were found in Bobylev and Bajkova (2013) from masers,
where the method of allowance for these perturbations directly in
the components $V_r,V_l$, and $V_b$  is described in detail.

We can see from the figure that waves are clearly visible in the
immediate solar neighborhood on the two left panels ($u=f(x)$ and
$v=f(x)$). Thus, there is no doubt that the periodic velocity
perturbations manifesting themselves in Fig.\ref{fig2} (panel (a))
and Fig.\ref{fig3} are related to the influence of the Galactic
spiral density wave.

%%%%%%%%%%%%%%%%%%%%%%%%%%%%%%%%%%%%%%%% FIG.4:
%%%%%%%%%%%%%%%%%%%%%%%%%%%%%%%%%%%%%%%%%%%%%%%
%%%%%%%%%%%%%%%%%%%%%%%%%%%%%%%%%%%%%%%% FIG.5:
%%%%%%%%%%%%%%%%%%%%%%%%%%%%%%%%%%%%%%%%%%%%%%%
%%%%%%%%%%%%%%%%%%%%%%%%%%%%%%%%%%%%%%%% FIG.6:
 \begin{figure}[t]
 {\begin{center}
 \includegraphics[width=100mm]{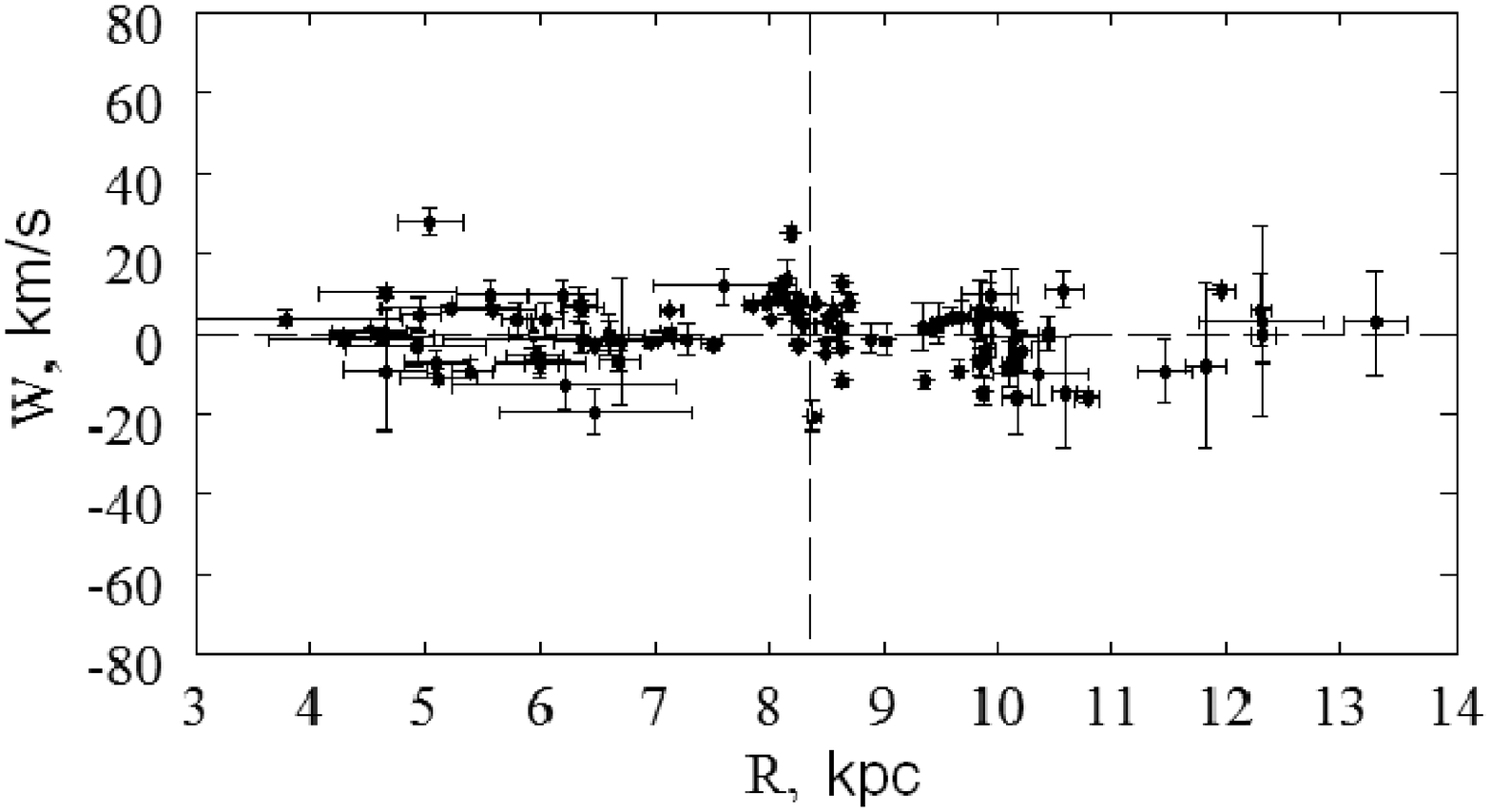}
 \caption{
Velocity $w$ versus Galactocentric distance $R$; the vertical line
marks the Sun’s position found. }
 \label{fig6}
 \end{center} }
 \end{figure}
%%%%%%%%%%%%%%%%%%%%%%%%%%%%%%%%%%%%%%%%%%%%%%%

If the periodicity in the velocities $w$ as a function of $x$
(Fig.\ref{fig2}) is related to the spiral structure, then the
dependence of $w$ on $R$ must be more interesting. This dependence
is presented in Fig.\ref{fig6}. We can see from the figure that
the picture is blurred in the inner region of the Galaxy, but the
masers in the Local arm and the Perseus arm ($R\approx10$ kpc)
have significant perturbations. About 30 masers belong to the
Local arm. The velocity distribution shown in Fig.\ref{fig6} comes
as no surprise for the Gould Belt. Indeed, if the Gould Belt stars
are subjected to intrinsic expansion away from the center
$l\approx180^\circ$ at an angle to the Galactic plane of
$20^\circ$ (Bobylev 2006), then there will be projections of these
velocities onto the $z$ axis. The projections of these velocities
will be positive for the stars of the Scorpius–Centaurus
association ($R<R_0$) and negative for the stars of the Orion
association ($R>R_0$), exactly as in the figure. Remarkably,
removing the Gould Belt objects ($r<0.5$ kpc) from the sample does
not change the picture. Thus, the entire Local arm shows a
velocity perturbation in $z,$ for which a single mechanism, for
example, the density wave, can be responsible.

Finally, note the paper by Branham (2014), who analyzed the
three-dimensional kinematics of more than 6000 OB stars within
$r<3$ kpc of the Sun. He used a method based on the generalized
Oort formulas, with two more terms having been added to Eqs.
(\ref{Ogor-01})--(\ref{Ogor-03}): $\partial w/\partial z$ and
$\partial^2 w/\partial^2 z$. As can be seen on panel (i) in Fig.
\ref{fig2}, even the linear trend ($\partial w/\partial z$) is
invisible. Therefore, adding these two terms when studying the
masers seems inappropriate so far.

\section*{DISCUSSION}
Using the generalized Oort formulas, Branham (2014) estimated the
solar Galactocentric distance from OB stars to be
$R_0=6.72\pm0.39$ kpc. This estimate differs noticeably from the
estimates of other authors.

Bobylev and Bajkova (2014) applied a method based on Bottlinger’s
formulas (\ref{Bottl-0234}) but without expanding Eq. (\ref{RR})
in a series, because R was calculated directly from this formula
using the distances r to stars. The following solution was
obtained from data on 73 masers:
 $u_\odot= 7.81\pm0.63~$km s$^{-1}$,
 $v_\odot=17.47\pm0.33~$km s$^{-1}$,
 $w_\odot= 7.73\pm0.23~$km s$^{-1}$,
 $\omega_0  =-28.86\pm0.45$~km s$^{-1}$ kpc$^{-1}$,
 $\omega'_0 = 3.96\pm0.09$~km s$^{-1}$ kpc$^{-2}$,
 $\omega''_0=-0.790\pm0.027$~km s$^{-1}$ kpc$^{-3}$,
$R_0=8.3\pm0.2$~kpc. In this case, the Oort constants are
$A=16.49\pm0.60$ km s$^{-1}$ kpc$^{-1}$ and $B=-12.37\pm1.12$ km
s$^{-1}$ kpc$^{-1}$. It can be seen that all parameters are
determined in this case with smaller errors than those in
solutions (\ref{rez-3}), (\ref{rez-4}), or (\ref{rez-55}), but the
$R_0$ estimation methods differ.

Our new estimate of $R_0=8.3\pm0.4$ kpc obtained here is
consistent with the results of other authors found by applying
different methods based on different data. Some of the results
were pointed out in the Introduction. We can point out several
more interesting works. Having analyzed the orbits of stars moving
around the massive black hole at the Galactic center (the
dynamical parallax method), Gillessen et al. (2009) obtained an
estimate of $R_0=8.33\pm0.35$ kpc. According to VLBI measurements,
the radio source Sqr~A* has a proper motion relative to
extragalactic sources of $6.379\pm0.026$ mas yr$^{-1}$ (Reid and
Brunthaler 2004), using which Sch$\ddot{o}$nrich (2012) found R0 =
$R_0=8.27\pm0.29$ kpc. Two H$^2$O maser sources, Sgr B2N and Sgr
B2M, are located in the immediate vicinity of the Galactic center,
where the radio source Sqr~A* lies. Based on their direct
trigonometric VLBI measurements, Reid et al. (2009b) obtained an
estimate of $R_0=7.9^{+0.8}_{-0.7}$ kpc. Francis and Anderson
(2014) gave a summary of 135 publications devoted to the R0
determination from 1918 to 2013. They concluded that the results
obtained after 2000 give a mean value of $R_0$ close to 8.0 kpc.

The detected oscillations of the velocities $w$ as a function of
$x$ (panel (g) in Fig.\ref{fig2}) or $R$ (Fig.\ref{fig6}) show
that beyond-the-model (with respect to the linear
Ogorodnikov.Milne model) harmonics in the motion of stars
(Vityazev and Tsvetkov 2014) can actually manifest themselves in
different planes, but they have the pattern of local
perturbations. Obviously, if the sample radius is small, then the
amplitude of such local harmonics can be significant. Indeed, if a
linear trend is drawn in Fig.\ref{fig6} through the Local arm
masers, then a large coefficient $\partial w/\partial R\approx-22$
km s$^{-1}$ kpc$^{-1}$ determined with a small error is obtained.

The perturbations only in the radial and tangential velocities are
considered in most models of the Galactic spiral structure.
Interestingly, large-scale vertical velocity perturbations have
been found recently in the Galactic disk based on data from
several experiments, such as SEGUE (Widrow et al. 2012), RAVE
(Williams et al. 2013), and LAMOST (Carlin et al. 2013). The
nonzero vertical velocities of objects are usually explained by
the action of some external factors, for example, by the passage
of a dwarf galaxy or clouds of dark matter through the Galactic
disk. However, such perturbations can also be explained without
invoking the action of external forces. Small velocity
perturbations in the vertical direction can take place within the
density wave theory. For example, Fridman (2007) pointed out the
possibility of such perturbations. The numerical simulations
performed, for example, by Faure et al. (2014) or Debattista
(2014) have shown that the propagation of a spiral density wave in
the Galactic disk can give rise to vertical oscillations with an
amplitude of 10–-20 km s$^{-1}$.

At present, the observations of $\approx 400$ masers are being
carried out within the framework of the BeSSeL project (Brunthaler
et al. 2011) with the goal of determining their trigonometric
parallaxes. Therefore, it will soon be possible to study in more
detail the above subtle kinematic effects based on a large sample
of sources.

\section*{CONCLUSIONS}
We gathered information about the coordinates, line-of-sight
velocities, proper motions, and trigonometric parallaxes of
Galactic masers from published data. Based on these data, we
tested a nonlinear model of Galactic rotation using generalized
Oort formulas. This model is interesting in that, apart from the
rotation parameters $\omega_0,$ $\omega'_0,$ and $\omega''_0$  it
allows the Galactocentric distance of the Sun $R_0$ to be easily
estimated.

Using a sample of masers with parallax errors
$\varepsilon_\pi/\pi<10\%$, we found the Galactic rotation
parameters and obtained a kinematic estimate of the solar
Galactocentric distance, $R_0=8.3\pm0.4$ kpc.

Analysis of the three-dimensional kinematics of 106 masers showed
significant nonlinearities to be present in the $xy$ plane
(rotation around the Galactic z axis) due to the peculiarities of
the Galactic rotation curve (the existence of high-order
derivatives in the expansion of the angular velocity). No distinct
linear trends were found in the other two planes, $xz$ and $yz$.
The presence of a wave in the velocities $w$ as a function of
coordinate $x$ or $R$ with a wavelength of $\approx 3$ kpc and an
amplitude of 10 km s$^{-1}$ is of considerable interest. This wave
is probably associated with the Galactic spiral density wave.

\section*{ACKNOWLEDGMENTS}
We are grateful to the referee for the useful remarks that
contributed to an improvement of the paper. This work was
supported by the “Nonstationary Phenomena in Objects of the
Universe” Program P-21 of the Presidium of the Russian Academy of
Sciences.

\bigskip

\bigskip

{\noindent\bf\Large REFERENCES}

\bigskip

\bigskip
{

1.Y.Asaki,H. Imai,A. M. Sobolev, and S. Yu. Parfenov,
arXiv:1404.6947 (2014).

2. V. V. Bobylev, Astron. Lett. 32, 816 (2006).

3. V. V. Bobylev, A. T. Bajkova, and S. V. Lebedeva, Astron. Lett.
33, 720 (2007).

4. V. V. Bobylev and A. T. Bajkova, Astron. Lett. 39, 809 (2013).

5. V. V. Bobylev, Astron. Lett. 39, 819 (2013).

6. V. V. Bobylev and A. T. Bajkova, Astron. Lett. 40, 389 (2014).

7. R. L. Branham, Astrophys. Space Sci. 999, 242 (2014).

8. A. Brunthaler, M. J. Reid, K. M. Menten, X.- W. Zheng, A.
Bartkiewicz, Y. K. Choi, T. Dame, K. Hachisuka, K. Immer, G.
Moellenbrock, et al., Astron. Nachr. 332, 461 (2011).

9. R. A. Burns, Y. Yamaguchi, T. Handa, T. Omodaka, T. Nagayama,
A. Nakagawa, M. Hayashi, T. Kamezaki, J. O. Chibueze, et al.,
arXiv:1404.5506 (2014).

10. J. L. Carlin, J. DeLaunay, H. J. Newberget, L. Deng, D. Gole,
K. Grabowski, G. Jin, C. Liu, X. Liu, et al., Astrophys. J. 777,
L5 (2013).

11. J. O. Chibueze, H. Sakanoue, T. Nagayama, T. Omodaka, T.
Handa, T. Kamezaki, R. Burns, H. Kobayashi, H. Nakanishi, M.
Honma, et al., arXiv:1406.277 (2014).

12. D. P. Clemens, Astrophys. J. 295, 422 (1985).

13. V. Debattista, Mon. Not. R. Astron. Soc. 443, L1 (2014).

14. S.Dzib, L. Loinard,A. J. Mioduszewski,A. F. Boden, L. F.
Rodriguez, and R. M. Torres, Astrophys. J. 718, 610 (2010).

15. C. Faure, A. Siebert, and B. Famaey, Mon. Not. R. Astron. Soc.
440, 2564 (2014).

16. T. Foster and B. Cooper, ASP Conf. Ser. 438, 16 (2010).

17. C. Francis and E. Anderson, Mon. Not. R. Astron. Soc. 441,
1105 (2014).

18. A.M. Fridman, Phys. Usp. 50, 115 (2007).

19. S. Gillessen, F. Eisenhauer, S. Trippe, T. Alexander, R.
Genzel, F. Martins, and T. Ott, Astroph. J. 692, 1075 (2009).

20. T. Hirota, T. Bushimata, Y. K. Choi, M. Honma, H. Imai, K.
Iwadate, T. Jike, S. Kameno, O. Kameya, R. Kamohara, et al., Publ.
Astron. Soc. Jpn. 59, 897 (2007).

21. T. Hirota, T. Bushimata, Y. K. Choi, M. Honma, H. Imai, K.
Iwadate, T. Jike, O. Kameya, R. Kamohara, et al., Publ. Astron.
Soc. Jpn. 60, 37 (2008).

22. M. Honma, T. Nagayama, K. Ando, T. Bushimata, Y. K. Choi, T.
Handa, T. Hirota, H. Imai, T. Jike, et al., Publ. Astron. Soc.
Jpn. 64, 136 (2012).

23. H. Imai, N. Sakai, H. Nakanishi, H. Sakanoue, M. Honma, and
T.Miyaji, Publ. Astron. Soc. Jpn. 64, 142 (2012).

24. M. K. Kim, T. Hirota, M. Honma, H. Kobayashi, T. Bushimata, Y.
K. Choi, H. Imai, K. Iwadate, T. Jike, S. Kameno, et al., Publ.
Astron. Soc. Jpn. 60, 991 (2008).

25. K. Kusuno, Y. Asaki, H. Imai, and T. Oyama, Astrophys. J. 774,
107 (2013).

26. E. S. Levine, C. Heiles, and L. Blitz, Astrophys. J. 679, 1288
(2008).

27. C. C. Lin and F. H. Shu, Astrophys. J. 140, 646 (1964).

28. L. Loinard, R. M. Torres, A. J. Mioduszewski, L. F. Rodriguez,
R. A. Gonzalez-Lopezlira, R. Lachaume, V. Vazquez, and E.
Gonzalez, Astrophys. J. 671, 546 (2007).

29. L. Loinard, R. M. Torres, A. J. Mioduszewski, and L. F.
Rodriguez, Astrophys. J. 675, 29 (2008).

30. N. M. McClure-Griffiths, and J. M. Dickey,Astroph. J. 671, 427
(2007).

31. A. M. Mel’nik, and A. K. Dambis, Mon. Not. R. Astron. Soc.
400, 518 (2009).

32. K. F. Ogorodnikov, Dinamika zvezdnykh sistem (M.: Fizmatgiz,
1965).

33. M. E. Popova and A. V. Loktin, Astron. Lett. 31, 663 (2005).

34. M. J. Reid, Annu. Rev. Astron. Astrophys. 31, 345 (1993).

35. M. Reid and A. Brunthaler, Astroph. J. 616, 872 (2004).

36. M. J. Reid, K. M. Menten, X. W. Zheng, A. Brunthaler, L.
Moscadelli, Y. Xu, B. Zhang, M. Sato, M. Honma, T. Hirota, et al.,
Astrophys. J. 700, 137 (2009a).

37. M. Reid, K. M. Menten, X. W. Zheng, A. Brunthaler, and Y. Xu,
Astrophys. J. 705, 1548 (2009b).

38. M. J. Reid, J. E. McClintock, R. Narayan, L. Gou, R. A.
Remillard, and J. A. Orosz, Astrophys. J. 742, 83 (2011).

39. M. J. Reid, K. M. Menten, A. Brunthaler, X.W. Zheng, T.M.
Dame, Y. Xu, Y.Wu, B. Zhang, A. Sanna, M. Sato, et al., Astrophys.
J. 783, 130 (2014).

40. K. L. J. Rygl, A. Brunthaler, M. J. Reid, K. M. Menten, H. J.
van Langevelde, and Y. Xu, Astron. Astrophys. 511, A2 (2010).

41. R. Sch . onrich, Mon. Not. R. Astron. Soc. 427, 274 (2012).

42. R. M. Torres, L. Loinard, A. J. Mioduszewski, and L. F.
Rodriguez, Astrophys. J. 671, 1813 (2007).

43. R. M. Torres, L. Loinard, A. J. Mioduszewski, and L. F.
Rodriguez, Astrophys. J. 698, 242 (2009).

44. R. M. Torres, L. Loinard, A. J. Mioduszewski, A. F. Boden, R.
Franco-Hernandez, W. H. T. Vlemmings, and L. F. Rodriguez,
Astrophys. J. 747, 18 (2012).

45. V. V. Vityazev and A. S. Tsvetkov, Mon. Not. R. Astron. Soc.
442, 1249 (2014).

46. L. M. Widrow, S. Gardner, B. Yanny, S. Dodelson, andH.-Yu.
Chen, Astrophys. J. 750, L41 (2012).

47. M. E. K. Williams, M. Steinmetz, J. Binney, A. Siebert, H.
Enke, B. Famaey, I. Minchev, R. S. de Jong, C. Boeche, K. C.
Freeman, et al.,Mon. Not. R. Astron. Soc. 436, 101 (2013).

48. M. V. Zabolotskikh, A. S. Rastorguev, and A. K. Dambis,
Astron. Lett. 28, 454 (2002).

49. J. Ziolkowski, Mon. Not. R. Astron. Soc. 358, 851 (2005).

 }

\end{document}